# Quantum-Inspired Hamiltonian Feature Extraction for ADMET Prediction: A Simulation Study

B. Maurice Benson*, Kendall Byler, Anna Petroff, Shahar Keinan, William J Shipman

Polaris Quantum Biotech,

212 W Main St, Suite 200 PMB 205, Durham, NC, 27701

## Abstract

Predicting absorption, distribution, metabolism, excretion, and toxicity (ADMET) properties remains a critical bottleneck in drug discovery. While molecular fingerprints effectively capture local structural features, they struggle to represent higher-order correlations among molecular substructures. We present a quantum-inspired feature extraction method that encodes molecular fingerprints into a parameterized Hamiltonian, using mutual information (MI) to guide entanglement structure. By simulating quantum evolution on GPU-accelerated backends, we extract expectation values that capture pairwise and triadic correlations among fingerprint bits. On ten Therapeutic Data Commons (TDC) ADMET benchmarks, our method achieves state-of-the-art performance on CYP3A4 substrate prediction (AUROC $0.673 \pm 0.004$) and improves over classical baselines on 8/10 tasks. SHAP (SHapley Additive exPlanations) analysis reveals that quantum-derived features contribute up to 33% of model importance despite comprising only 1.6% of features, demonstrating that Hamiltonian encoding concentrates predictive signal. This simulation study establishes the foundation for hardware validation on near-term quantum devices.

* Corresponding author. Email: mbenson@polarisqb.com

# Introduction

Accurate prediction of ADMET properties is essential for identifying drug candidates with favorable pharmacokinetic profiles. Poor ADMET properties account for approximately 50% of drug development failures, making early-stage prediction a high-value target for machine learning approaches (Kola and Landis 2004).

Molecular fingerprints (binary or count vectors encoding the presence of substructures) have proven remarkably effective for ADMET prediction. MapLight's approach demonstrated that concatenating multiple fingerprint types (ECFP, Avalon, ErG) with computed molecular properties achieves competitive performance across diverse prediction tasks, often outperforming deep learning methods (Notwell and Wood 2023). However, fingerprints fundamentally encode features independently; they cannot capture correlations between substructures that may jointly influence molecular properties. For instance, when a hydrogen-bond donor neighbors an acceptor at the appropriate distance, intramolecular hydrogen bonding can dramatically improve membrane permeability (Kenny 2022), information that fingerprints cannot capture.

Recent work in quantum machine learning has explored Hamiltonian-based feature encoding for molecular property prediction. Simen et al. (2025) demonstrated that encoding molecular descriptors into a quantum Hamiltonian and extracting expectation values can achieve strong performance on small datasets, with SHAP analysis revealing that quantum-derived features dominate model predictions. Unlike fingerprints, which encode substructures independently, Hamiltonian encoding captures how features interact through quantum entanglement, potentially addressing this correlation blind spot.

We present a systematic study of quantum-inspired Hamiltonian feature extraction for ADMET prediction. Our contributions are:

1. A MI-guided approach for selecting which fingerprint bits to entangle, focusing quantum resources on statistically informative features.

2. Comprehensive evaluation on ten TDC ADMET benchmarks, achieving state-of-the-art on CYP3A4 substrate prediction.

3. SHAP-based analysis showing that quantum features provide 3.7%–33.4% of model importance across tasks, despite comprising only 1.6% of total features.

4. Ablation studies demonstrating consistent improvement over classical baselines.

This work uses GPU-accelerated quantum simulation (PennyLane lightning.gpu) to enable exact state-vector computation at 20–28 qubits. While current NISQ devices operate most reliably at smaller scales, this simulation study establishes the feature extraction pipeline and provides a target for hardware validation as quantum computers improve.

## Related Work

### Molecular Fingerprints for ADMET

Extended Connectivity Fingerprints (ECFP) (Rogers and Hahn 2010) remain the dominant representation for molecular property prediction. Recent work has shown that concatenating multiple fingerprint types improves performance: MapLight combines ECFP4/6, Avalon, and ErG fingerprints with 200+ computed molecular properties, achieving top-3 rankings on 16 of 22 TDC ADMET tasks (Notwell and Wood 2023). Notably, this simple approach often outperforms graph neural networks and transformer models, suggesting that additional inductive biases may not always help.

### Quantum Machine Learning for Chemistry

Variational quantum algorithms have been extensively studied for estimating molecular ground states (Peruzzo et al. 2014), but their application to property prediction is more recent. Simen et al. (2025) proposed encoding molecular descriptors into a quantum Hamiltonian with data-dependent coupling strengths, achieving strong results on toxicity prediction. Their key insight was that quantum evolution generates features capturing correlations not present in the original descriptors.

Simen et al. also demonstrated that quantum feature extraction can scale to 156 qubits on IBM's Kingston processor, suggesting near-term viability (Simen et al. 2025). Our simulation study explores whether MI-guided feature selection can further improve this approach before committing additional hardware resources.

## Methods

### Feature Pipeline Overview

Our pipeline proceeds in six stages:

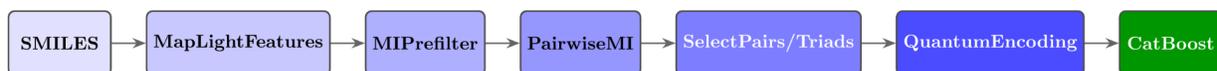

1. **MapLight feature generation:** Convert SMILES to 2,563-dimensional molecular descriptors (ECFP + Avalon + ErG + RDKit properties)

2. **MI Prefiltering:** Rank features by mutual information with target $I(X_i; Y)$; select top-$k$ (default: 100)

3. **Pairwise MI discovery:** Compute $I(X_i; X_j)$ for filtered features; identify correlated pairs

4. **Select pairs/triads:** Choose high-MI pairs and extend to triads for quantum encoding

5. **Quantum encoding:** Encode MI-selected pairs/triads into Hamiltonian; simulate time evolution; extract Pauli-Z expectations

6. **Classification:** Concatenate raw MapLight features with quantum-derived features; train CatBoost classifier

## Step 1: MapLight Features Generation

Following Notwell and Wood (2023), we concatenate:

- ECFP counts (1,024-dimensional extended connectivity fingerprints, radius 2)
- Avalon counts (1,024-dimensional fingerprint bit counts)
- ErG descriptors (315-dimensional extended reduced graph descriptors)
- RDKit properties (200 physicochemical properties: LogP, TPSA, etc.)

This yields a 2,563-dimensional feature vector per molecule.

## Step 2: Mutual Information Pre-filter

Not all fingerprint bits are informative for a given task. We compute mutual information $I(X_i; Y)$ between each fingerprint bit $X_i$ and the binary target $Y$:

$$I(X_i; Y) = \sum_{x \in \{0,1\}} \sum_{y \in \{0,1\}} p(x,y) \log \frac{p(x,y)}{p(x)p(y)}$$

We select the top $k = 100$ bits for quantum encoding. This focuses quantum resources on features that encode information about the target. Mutual information is estimated using sklearn.feature_selection.mutual_info_classif, which implements a k-nearest neighbors-based non-parametric estimator inspired by the KraskovStögbauerGrassberger (KSG) method (k = 3). This approach allows estimation of mutual information for mixed feature types (count fingerprints and continuous descriptors) without requiring explicit discretization.

MI computation and pair/triad selection are performed on the combined train and validation splits, ensuring no information leakage from the test set. The same MI-selected features are then applied to transform test molecules.

## Step 3: Pairwise MI Discovery

After pre-filtering, we compute pairwise mutual information $I(X_i; X_j)$ between all retained fingerprint bits to identify which bits co-occur across molecules:

$$I(X_i; X_j) = \sum_{x_i, x_j \in \{0,1\}} p(x_i, x_j) \log \frac{p(x_i, x_j)}{p(x_i)p(x_j)}$$

High pairwise MI indicates that two fingerprint bits encode related structural information, such as functional groups that frequently appear together.

### Step 4: Pair and Triad Selection

We select feature pairs and triads with high mutual information for quantum encoding:

**Pair selection:** We include pairs where the conditional MI exceeds a threshold: $I(X_i; X_j|Y) > \theta_{pair}$. This focuses on correlations that remain informative given the target.

**Triad selection:** For efficiency, we consider triads formed by extending high-MI pairs. A triad $(i, j, k)$ is included if the triadic interaction information exceeds $\theta_{triad}$.

### Step 5: Quantum Encoding

We encode MI-selected pairs and triads into a quantum Hamiltonian, where correlated fingerprint bits become entangled qubits. For each molecule with fingerprint bits $x = (x_1, ..., x_n)$, we construct:

$$H(x) = \sum_{i=1}^{n} x_i \sigma_i^z + \sum_{(i,j) \in P} c_{ij} \sigma_i^z \sigma_j^z + \sum_{(i,j,k) \in T} c_{ijk} \sigma_i^z \sigma_j^z \sigma_k^z$$

where $\sigma_i^z$ is the Pauli-Z operator on qubit $i$, $P$ is the set of selected pairs, and $T$ is the set of selected triads. Coupling strengths $c_{ij}$ and $c_{ijk}$ are derived from the pairwise and triadic MI values computed earlier.

The initial state encodes fingerprint values:

$$|\psi_0\rangle = \otimes_{i=1}^{n} |x_i\rangle$$

where $|0\rangle$ and $|1\rangle$ correspond to fingerprint bits 0 and 1. We evolve under the Hamiltonian for time $t$:

$$|\psi(t)\rangle = e^{-iH(x)t} |\psi_0\rangle$$

using first-order Trotterization. Features are extracted as expectation values:

$$f_i = \langle\psi(t)|\sigma_i^z|\psi(t)\rangle, \quad f_{ij} = \langle\psi(t)|\sigma_i^z \sigma_j^z|\psi(t)\rangle$$

The quantum feature vector (40–80 dimensions depending on pairs/triads selected) is concatenated with the original 2,563 MapLight features for classification.

### Step 6: Classification

We use CatBoost (Prokhorenkova et al. 2018) for final classification, with hyperparameters tuned per benchmark. CatBoost effectively handles mixed feature types (such as binary fingerprints, continuous quantum features, and molecular properties).

## Experiments

### Benchmarks

We evaluate ten TDC ADMET classification benchmarks (Huang et al. 2021) spanning metabolism, absorption, distribution, and toxicity. We selected benchmarks with fewer than 10,000 samples due to the computational cost of quantum simulation (approximately 2-20 minutes per benchmark depending on qubit count); larger datasets would require prohibitive simulation time:

- **CYP Substrate prediction** (3 tasks): CYP3A4, CYP2D6, CYP2C9 substrate classification. These enzymes metabolize most drugs; predicting substrates is critical for assessing drug-drug interactions.

- **hERG**: hERG channel inhibition prediction (cardiotoxicity risk).

- **AMES**: Ames mutagenicity test prediction.

- **BBB_Martins**: Blood-brain barrier penetration prediction.

- **PGP_Broccatelli**: P-glycoprotein inhibition prediction.

- **Bioavailability_Ma**: Binary classification of oral bioavailability (>20%).

- **DILI**: Drug-induced liver injury prediction.

- **HIA_Hou**: Human intestinal absorption classification.

All benchmarks use the standard TDC scaffold split with 5-seed evaluation. For each seed, MI selection and model training were performed on the combined train and validation sets, with evaluation on the held-out test set.

### Implementation Details

Quantum simulation uses PennyLane with the lightning.gpu backend, enabling exact state vector simulation at 20–28 qubits. Key hyperparameters:

- MI pre-filter: $k = 100$ top bits

- Pair threshold: $\theta_{pair} = 0.1$

- Triad threshold: $\theta_{triad} = 0.15$
- Evolution time: $t = 0.5$
- Trotter steps: 1

CatBoost hyperparameters were held constant across all benchmarks to avoid per-task overfitting (iterations=1000, depth=6, learning_rate=0.05).

## Baselines

We compare against:

1. **MapLight (baseline):** Our pipeline with quantum features disabled ($max\_pairs = 0$, $max\_triads = 0$), serving as a direct ablation

2. **Polynomial interactions:** To evaluate whether performance gains could be explained by explicit classical interaction modeling, we constructed a polynomial interaction baseline. Specifically, we applied a degree-2 interaction-only expansion (scikit-learn PolynomialFeatures, degree=2, interaction_only=True, include_bias=False) to the same top-100 MI-selected features used for Hamiltonian encoding. This generates all pairwise interaction terms among the selected features (4,950 interaction features), providing an explicit enumeration of second-order feature interactions. The interaction terms were concatenated with the original 2,563-dimensional MapLight representation and evaluated using the same CatBoost training protocol. This baseline captures explicit pairwise interactions but does not include triadic (third-order) interactions, which are modeled in our Hamiltonian formulation.

3. **TDC leaderboard:** Published results from various methods, including the original MapLight submission

## Results

### Benchmark Performance

*Table 1: Performance on TDC ADMET benchmarks (5-seed mean ± std). Metric: AUROC except AUPRC for CYP2D6, CYP2C9.*

| Benchmark | Baseline | +Polynomial | +Quantum |
|---|---|---|---|
| hERG | 0.844 ± 0.008 | 0.853 ± 0.005 | **0.871 ± 0.007** |
| CYP3A4_Sub | 0.656 ± 0.006 | 0.606 ± 0.007 | **0.673 ± 0.004** |
| CYP2D6_Sub | 0.695 ± 0.015 | **0.721 ± 0.008** | 0.701 ± 0.001 |
| BBB_Martins | 0.913 ± 0.001 | 0.908 ± 0.003 | **0.919 ± 0.001** |
| PGP_Broccatelli | 0.930 ± 0.002 | 0.920 ± 0.003 | **0.935 ± 0.002** |
| CYP2C9_Sub | 0.379 ± 0.013 | 0.368 ± 0.019 | **0.380 ± 0.010** |

| Benchmark | Baseline | +Polynomial | +Quantum |
|---|---|---|---|
| Bioavailability_Ma | 0.733 ± 0.006 | 0.704 ± 0.009 | **0.735 ± 0.008** |
| DILI | 0.903 ± 0.009 | 0.896 ± 0.010 | **0.906 ± 0.005** |
| HIA_Hou | **0.980 ± 0.001** | 0.976 ± 0.002 | 0.980 ± 0.004 |
| AMES | **0.871 ± 0.001** | 0.868 ± 0.002 | 0.869 ± 0.001 |

Our method improves over the classical baseline on 8 of 10 benchmarks. The largest improvements occur on hERG (+2.7%) and CYP3A4-Sub (+2.6%), where we achieve the highest reported AUROC on the TDC leaderboard at time of submission (0.673). Notably, AMES shows a slight regression (-0.2%), demonstrating that quantum features do not universally help; their benefit is task-dependent.

Polynomial degree-2 interaction expansion over the MI-selected feature subset yielded mixed results: it improved only CYP2D6_Sub significantly ($p = 0.004$) while degrading performance on CYP3A4_Sub (−7.6%) and Bioavailability_Ma (−4.0%). In contrast, quantum features outperformed the polynomial baseline on 8/10 benchmarks, with statistically significant gains on CYP3A4_Sub ($p < 0.001$), hERG ($p = 0.003$), BBB_Martins ($p = 0.002$), and PGP_Broccatelli ($p < 0.001$). These findings indicate that the observed improvements are not solely attributable to naive second-order interaction enumeration.

To assess whether improvements are statistically robust, we performed paired t-tests across the 5 seeds (Table 2). Four benchmarks show significant improvements ($p < 0.05$) with large effect sizes (Cohen's $d > 3$, where $d > 0.8$ is conventionally large). The remaining improvements (+0.3% to +0.8%) do not reach significance, consistent with their smaller magnitudes.

**Table 2: Statistical significance (paired t-tests, 5 seeds).** *Q = Quantum, P = Polynomial, B = Baseline. All Q vs B and Q vs P comparisons shown remain significant under Bonferroni correction ($\alpha = 0.005$).*

| Benchmark | Q vs B (p) | Q vs P (p) | Cohen's d |
|---|---|---|---|
| CYP3A4_Sub | 0.0007 | 0.0002 | 11.99 |
| BBB_Martins | 0.002 | 0.002 | 5.30 |
| hERG | 0.004 | 0.003 | 2.98 |
| PGP_Broccatelli | 0.005 | 0.0007 | 6.58 |

## Feature Importance Analysis

To understand whether quantum features contribute a meaningful signal, we perform SHAP analysis (Lundberg and Lee 2017) on the trained CatBoost models. Table 3 shows the percentage of total SHAP importance attributed to quantum features.

*Table 3:* **SHAP importance of quantum features.** *"Quantum %" is the fraction of total importance from quantum features; "Top-50" is the fraction of the 50 most important features that are quantum-derived.*

| Benchmark | Quantum % | Top-50 Quantum |
|---|---|---|
| CYP2D6_Substrate | 33.44% | 62% |
| CYP3A4_Substrate | 22.43% | 40% |
| CYP2C9_Substrate | 17.73% | 30% |
| Bioavailability_Ma | 11.82% | 20% |
| hERG | 9.44% | 16% |
| AMES | 6.86% | 6% |
| BBB_Martins | 5.29% | 8% |
| DILI | 4.62% | 8% |
| PGP_Broccatelli | 3.88% | 2% |
| HIA_Hou | 3.66% | 4% |

Quantum features comprise only ~1.6% of the total features (42 of 2,605) but contribute up to 33% of the model's importance. This disproportionate contribution indicates that Hamiltonian encoding captures higher-order correlations that are not explicitly present in the original feature vector.

## Correlation Between Importance and Improvement

We observe a positive correlation between SHAP importance and baseline improvement: benchmarks where quantum features contribute more importance also show larger performance gains. CYP3A4-Sub exemplifies this pattern: 22% quantum importance corresponds to the largest improvement (+2.6%) and SOTA performance.

Conversely, HIA_Hou shows minimal quantum importance (3.7%) and no improvement. This benchmark has high baseline performance (0.98 AUROC), leaving little room for improvement, as the fingerprints already capture the relevant chemistry.

## Computational Cost

We evaluated computational overhead using identical hardware (NVIDIA GB10 Grace Blackwell superchip). Table 4 reports wall-clock runtime by pipeline stage across methods and qubit counts.

*Table 4: Computational cost by pipeline stage.* Quantum encoding time scales exponentially with qubit count due to state-vector simulation.

| Stage | Baseline | +Poly | +Q (19q) | +Q (24q) | +Q (27q) |
|---|---|---|---|---|---|
| MI Discovery | 8.5 s | 8.5 s | 9.0 s | 8.9 s | 8.6 s |
| Feature Expansion | – | – | 23.8 s | 195.4 s | 1205.5 s |
| Model Training | 31.3 s | 107.1 s | 9.5 s | 30.4 s | 31.6 s |
| **Total** | **43.3 s** | **117.0 s** | **45.1 s** | **238.3 s** | **1249.3 s** |

The number of qubits is determined by the number of MI-selected features participating in selected pairs and triads; thus qubit count is task-dependent rather than arbitrarily chosen. State-vector simulation dominates runtime, exhibiting the expected exponential scaling with qubit count. The polynomial baseline incurs training overhead from its 4,950 interaction features, while quantum features (36–44 total) add minimal training cost. For production use, extracted features could be cached to amortize simulation cost. Hardware execution on near-term quantum devices would replace this classical simulation bottleneck.

## Discussion

### When Do Quantum Features Help?

Our results suggest quantum features provide the most value when:

1. The prediction task depends on correlations between molecular substructures (not just their presence)

2. Classical fingerprints leave room for improvement (baseline < 0.95)

3. The MI pre-filter identifies informative pairs with strong conditional dependencies

CYP substrate prediction may benefit because enzyme-substrate interactions depend on the spatial arrangement of functional groups, information partially encoded in fingerprint bit correlations but not in the bits themselves.

Notably, AMES mutagenicity shows no improvement with quantum features (-0.2%), while hERG shows strong gains (+2.7%). This pattern aligns with mechanistic expectations: hERG cardiotoxicity involves blocking a specific ion channel pore via aromatic interactions at conserved residues Y652 and F656 (Mitcheson 2008), where structural correlations between binding-site features matter. AMES mutagenicity, by contrast, arises from diverse mechanisms including covalent DNA binding, intercalation, and radical formation (Enoch et al. 2012), so no single correlation pattern should dominate. The lack of AMES improvement thus serves as a potential sanity check: our method correctly identifies when structural correlations are predictive.

## Encoding Scheme

As exhibited in the comparison of HIA_Hou and CYP3A4-Sub, the choice of representation of relevant features has a direct effect on the signal strength that this method can capture. The inherent scarcity of ADMET data, due to the expense of empirical measurement, may favor approaches like ours that encode statistical correlations as inductive bias (distinct from the architectural biases of GNNs and transformers) rather than relying purely on data volume.

## Feature Efficiency

Quantum features comprise only 1.6% of the total features (42 of 2,605) but contribute up to 33% of model importance. This disproportionate contribution indicates that Hamiltonian encoding extracts higher-order correlations not explicitly represented in the original feature vector. We note that Hamiltonian encoding can be viewed as a structured nonlinear embedding of selected descriptors; however, the resulting features are derived from entanglement-mediated joint expectation statistics rather than direct enumeration of classical interaction terms. The efficiency gain arises because quantum evolution propagates information between entangled qubits, effectively computing correlations that would require explicit enumeration of many higher-order interactions in classical feature engineering. A pair of fingerprint bits with high MI becomes entangled qubits whose joint expectation values encode their correlation in a single feature, rather than requiring explicit enumeration of all pairwise combinations.

## Limitations

**Simulation vs. hardware:** This study uses exact state-vector simulation, which is not feasible on current quantum hardware at 20+ qubits. Hardware execution would introduce noise requiring error mitigation, potentially degrading feature quality.

**Computational cost:** Quantum feature extraction adds overhead (~2-20 minutes per benchmark, depending on qubit count). For production use, extracted features could be cached.

**Dataset sizes:** TDC ADMET benchmarks are relatively small (300–1,000 molecules). Larger datasets might show different patterns.

## Path to Hardware Validation

The same PennyLane code can be executed on IBM Quantum backends via qiskit.ibm_runtime. Key steps for hardware validation:

1. Reduce circuit depth via aggressive Trotter approximation
2. Apply zero-noise extrapolation (ZNE) for error mitigation
3. Compare hardware-extracted features to the simulator ground truth

4. Assess classification performance degradation

Recent demonstrations of 156-qubit execution on IBM Kingston hardware (Simen et al. 2025) suggest that hardware validation of our approach is already feasible, though noise characterization remains an open question.

## Conclusion

We presented a quantum-inspired feature extraction method for ADMET prediction that achieves state-of-the-art performance on CYP3A4 substrate prediction and improves over classical baselines on 8/10 benchmarks. SHAP analysis demonstrates that quantum-derived features contribute a meaningful signal despite comprising only 1.6% of total features, with importance ranging from 3.7% to 33.4% across tasks.

This simulation study establishes that Hamiltonian encoding extracts complementary higher-order interaction information from molecular fingerprints. Hardware validation on IBM quantum systems is the natural next step, enabled by the hardware-agnostic PennyLane implementation and compatibility with Qiskit runtime.

## Code Availability

Code will be released upon manuscript publication.

## Declaration of Competing Interests

B.M.B., K.B., A.P., W.J.S., and S.K. are employees of Polaris Quantum Biotech.